\documentclass[aps,preprint,showpacs,amsmath,amssymb]{revtex4-1}
\usepackage{graphicx}
\usepackage{dcolumn}
\usepackage{bm}
\usepackage{hyperref}
\usepackage{xcolor}
\usepackage{soul}
\begin{document}

\title{Bound entanglement detection in  $4 \otimes 4$ systems via generalized Choi maps}
\author{Mazhar Ali \footnote{Email: mazharaliawan@yahoo.com or mazhar.ali@iu.edu.sa}}
\affiliation{Department of Electrical Engineering, Faculty of Engineering, Islamic University of Madinah, 
107 Madinah, Saudi Arabia}


\begin{abstract}
We construct a family of positive but not completely positive linear maps acting on $M_4(\mathcal{C})$, obtained as a natural extension of Kye’s indecomposable maps defined for $M_3(\mathcal{C})$. We rigorously investigate the conditions of positivity for these maps on general positive $4 \times 4$ matrices. We employ these conditions to detect a family of quantum states with both bound entangled states and free entangled states in $4 \otimes 4$ systems, a regime that remains less characterized compared to lower-dimensional cases. The proposed maps detect entanglement and reveal new structural features of the PPT entangled region in higher dimensions. Our results extend the applicability of positive-map–based entanglement detection and contribute to the systematic understanding of bound entanglement beyond the $3\otimes 3$ and $2 \otimes 4$ systems. In addition, we show that generalized Choi maps simply can not detect well known PPT entangled states for $2 \otimes 4$ systems.
\end{abstract}

\pacs{03.65.Db, 03.65.Ud, 03.67.-a}

\maketitle

\section{Introduction}\label{S-intro}

The characterization and detection of quantum entanglement is of utmost importance in quantum information theory \cite{Horodecki-RMP81,Guehne-PR474,Eisert-RMP82}, due to its potential utilization for upcoming quantum technologies \cite{Erhard-NRP-2020,Friis-2019}. Although the Peres–Horodecki criterion \cite{Peres-PRL77,Horodecki-PLA223} based on partial transposition provides a necessary and sufficient test for separability in low-dimensional systems such as $2 \otimes 2$ and $2 \otimes 3$, however, its sufficiency fails in higher dimensions. In these cases, there exist mixed states that are positive under partial transposition (PPT) yet entangled—commonly referred to as bound entangled states \cite{Horodecki-PLA232,Bennet-PRL82,Bruss-PRA61}. Such states cannot be distilled into pure entanglement by local operations and classical communication (LOCC) and thus represent a subtle and structurally rich form of quantum correlations. Numerous complementary approaches to entanglement detection have been developed. These include entanglement witnesses constructed via convex optimization \cite{Lewenstein-PRA62}, nonlinear entanglement criteria based on uncertainty relations \cite{Guehne-PRL92}, covariance matrices \cite{Toth-PRA72}, majorization relations \cite{Nielsen-PRL86}, realignment (or computable cross-norm) criteria \cite{Chen-QIC3, Rudolph-QIP4} and its variant \cite{Aggarwal-PRA109}, entanglement detection through local measurements or Bell-type inequalities \cite{Gisin-PLA154}, and partial transpose moments \cite{Schneeloch-PRR2, Elben-PRA99, Ketterer-PRL122}. In addition, numerous investigations have been done for entanglement detection in multipartite systems \cite{Jungnitsch-PRL106,Guehne-NJP12,Zhou-NPJ5, Xu-PRA107} and entanglement detection via moments of partial transposed matrix \cite{Zhang-QIP21,Zhang-AP534,Ali-QIP22,Wang-EPJ137}. Together, these methods form a rich hierarchy of separability tests, differing in their experimental accessibility and mathematical strength. The positive-map framework explored in this work complements these approaches by providing analytical insight into the geometric and algebraic structure of entanglement, especially in higher-dimensional PPT regions where other criteria may fail.

A powerful mathematical framework for studying entanglement is provided by the theory of positive but not completely positive linear maps. According to the Horodecki theorem \cite{Horodecki-PLA223}, a bipartite state is separable if and only if it remains positive under the action of all positive maps on one subsystem. Consequently, non-completely positive maps furnish entanglement witnesses and serve as essential tools for detecting states that escape simpler separability criteria. This is an extensive research area and lot of efforts are done to study positive maps which are not completely positive. Advances in this area have relied on the theory of positive but not completely positive maps, beginning with the pioneering constructions of Choi map \cite{Choi-LA10} and the development of indecomposable examples capable of detecting PPT entangled states \cite{Woronowicz-RMP10}. Notable contributions include the reduction map \cite{Horodecki-PRA59} together with the families proposed by Breuer \cite{Breuer-PRL97}, Hall \cite{Hall-JPA39}, Robertson \cite{Robertson-JLM32}, Hou \cite{Hou-JPA43} and the extensive investigations by Kye and co-workers on the structure and classification of positive linear maps \cite{Kye-work}.  
These developments collectively form the mathematical foundation for modern entanglement detection, upon which the present work extends the framework to higher-dimensional systems.

Despite substantial progress, the landscape of indecomposable positive maps in dimensions higher than three remains sparsely explored. In particular, generalizing known constructions while preserving positivity and indecomposability is highly nontrivial. In this work, we construct an explicit extension of Kye’s indecomposable maps to the four-dimensional case $M_4(\mathcal{C})$. We rigorously establish the conditions of positivity. We employ these conditions to detect bound entanglement in a family of quantum states in $4 \otimes 4$ states. This provides new analytical examples of positive maps which are not completely positive in higher dimensions and demonstrates the utility of generalized positive maps for probing the geometry of the PPT region. We also study Kye maps for $2 \otimes 4$ systems and show analytically that well known bound entangled states and all their variants, simply can not be detected by such maps. We need some other approach to construct positive maps to detect these states.  

In the present work we have focused on the characterization of entanglement using positive (but not completely positive) maps in a closed quantum system. In realistic scenarios, however, quantum systems inevitably interact with their surrounding environments, leading to open-system dynamics that may be either Markovian or non-Markovian. Environmental interactions generally induce decoherence and dissipation, which tend to degrade quantum correlations and may therefore reduce the effectiveness of entanglement detection schemes. In Markovian environments this degradation typically occurs monotonically, while non-Markovian dynamics can lead to memory effects that allow temporary revivals of quantum correlations. The influence of such environments on quantum correlations has been extensively investigated in the literature \cite{Nie-PRL101, Shen-PRA98,Breuer-PRL103}. A detailed study of how environmental noise affects the performance of the positive maps introduced here constitutes an interesting direction for future work.

It is also worth noting that recent experimental and educational studies have explored the classical optical emulation of certain quantum correlation phenomena. In particular, classical light fields have been used to emulate aspects of quantum state tomography, Bell-type tests, and biphoton spectroscopy in controllable laboratory settings. Such approaches provide experimentally accessible platforms for illustrating the structure of quantum correlations and related measurement protocols. For example, optical experiments have demonstrated the emulation of quantum state tomography and Bell-test procedures in undergraduate laboratory environments, as well as the reproduction of correlation signatures associated with entangled biphoton spectroscopy using classical light pulses. Although these systems do not generate genuine quantum entanglement, they can reproduce analogous correlation structures and measurement procedures, offering useful platforms for exploring entanglement-detection concepts. In this context, positive-map–based entanglement witnesses such as those studied in the present work could also be analyzed within similar emulation frameworks at the level of correlation measurements and tomographic reconstruction. Relevant demonstrations of such classical optical emulation have been reported in recent studies. \cite{Arbel-RO21, Ko-JPC14}

The remainder of this paper is organized as follows. In Sec. \ref{Sec:2}, we briefly review the necessary background on positive and completely positive maps. We present the construction of the extended Kye maps and analyze their structural properties. In Sec. \ref{Sec:M}, we apply these maps to a parameterized family of quantum states and prove the detection power of these positive maps to detect bound entanglement. In Sec. \ref{Sec:N}, we apply these maps to well known bound entangled states all their variants. Finally, Sec. \ref{Sec:Conc} summarizes our results and discusses possible directions for further generalizations.

\section{Separability for bipartite quantum systems} 
\label{Sec:2}

We define $\mathcal{H}_{AB} = \mathcal{H}_A \otimes \mathcal{H}_B$ as the bipartite finite dimensional Hilbert space. A mixed quantum 
state $\varrho \in \mathcal{H}_{AB}$ is a positive semidefinitive density matrix with unit trace. A bipartite density matrix is said to be separable if it can be written as 
\begin{eqnarray}
\varrho^{AB} = \sum_i^n \, p_i \,  \varrho_i^A \otimes \varrho_i^B \, 
\label{Eq:sep}
\end{eqnarray}
where $p_i \geq 0 $, $\sum_i \, p_i = 1$, and $\varrho_i^A$ ($\varrho_i^A$) is a state for subsystem $A$ ($B$). If a quantum state is not separable then it is entangled. However, it is not simple to use this definition to check whether a given quantum state can be written as convex combination of product states. Despite significant efforts and some partial results, it is still an open question to decide if a given quantum state is entangled or not \cite{Horodecki-RMP81,Guehne-PR474}. 

An important contribution on this problem of separabilty is to relate the issue with theory of positive maps \cite{Horodecki-PLA223}. A map $\Lambda$ is said to be positive if it maps a positive matrix $M$ (with positive eigenvalues) to another positive matrix $N =\Lambda (M)$. A positive map is called complete positive if $\mathcal{I}_A \otimes \Lambda_B (X)$ is positive, where  $X \in \mathcal{H}_{AB}$ is positive, otherwise $\Lambda$ is called positive but not completely positive map. Such maps are indecomposable and quite powerful for detection of entanglement. It was found that if $\Lambda$ is a positive map but not completely positive then for a separable state $\rho_{AB}$, the matrix $\mathcal{I} \otimes \Lambda (\rho_{AB})$ must have all positive eigenvalues. This condition is both necessary and sufficient for detection of entanglement \cite{Horodecki-PLA223}. This means that if a given quantum state $\sigma_{AB}$ is entangled then there must exist a positive map $\Lambda$, such that the matrix $\mathcal{I} \otimes \Lambda (\sigma_{AB})$ will have at least one negative eigenvalue. 
Therefore, it follows that the problem of detection of entanglement is to find the positive maps which are not completely positive. This issue is non-trivial because it is not easy to find the positive maps which are not completely positive. Even if we are able to find such maps, it is not clear whether they will detect a given quantum state. Hence the challenge is to look for 'quantum state specific positive maps' as other positive maps may not detect entanglement of a quantum given state.
  
Transposition is one such positive map, which is not completely positive and hence can detect some entangled states \cite{Peres-PRL77}. A necessary condition for separability is to check the partial transpose of the density matrix. If $(\varrho^{AB})^{T_B}$ is negative (having atleast one negative eigenvalue) then state $\varrho^{AB}$ is entangled. This condition is necessary and sufficient for $ 2 \otimes 2 $ and $ 2 \otimes 3$ quantum systems \cite{Horodecki-PLA223}. However, for higher dimensions of Hilbert space, there are quantum states having positive partial transpose (PPT) nevertheless entangled. In this work, we only focus on $2 \otimes 4$ and $4 \otimes 4$ quantum systems with Hilbert space having dimension $8$ and $16$, respectively. The $2 \otimes 4$ is the lowest dimension bipartite quantum system higher than $2 \otimes 3$ (dimension $6$). Already we observe that it is not an easy task to construct positive maps to detect PPT entangled states in this dimension. There are few known examples of PPT-entangled states for this system but it is not known how to construct the positive but not completely positive maps for them.  

In this work, we approach the problem from the perspective of some well known positive maps and study their range of entanglement detection. Kye’s indecomposable positive maps on $M_3(\mathcal{C})$ have played a central role in the study of bound entanglement. These maps provided some of the first systematic examples of analytically tractable, indecomposable maps capable of detecting PPT entangled states in $3 \otimes 3$ systems \cite{Kye-work}. Owing to their clear algebraic structure and close connection to unextendible product bases, Kye maps have become a benchmark for testing the strength of positive-map–based separability criteria. However, their applicability has been largely restricted to the three-dimensional case, leaving open the question of whether similar constructions can exist in higher dimensions. Extending these maps to $M_4(\mathcal{C})$ is  significant and provides new analytical tools for identifying bound entangled states. To describe it, we first define 
$| e_j\rangle$ with $j = 1, 2, 3, 4$, to be an orthonormal basis in $\mathcal{C}^4$. We define the elementary operators by 
$E_{ij} = |e_i\rangle\langle e_j|$ (we have $16$ such operators). In addition, we define two additional operators $F_{ij} = (E_{ii}+E_{jj})/\sqrt{2}$ and $G_{ij} = (E_{ii}-E_{jj})/\sqrt{2}$ \cite{Hou-JPA43}. The general positive Kye map originally defined for dimension $3$ systems \cite{Kye-work}, and extended for dimension $4$ systems, can be written as 
\begin{eqnarray}
\Phi[w,x,y,z] (X) = w \, \sum_{i} \, E_{ii} X E_{ii}^\dagger + x \, (E_{12} X E_{12}^\dagger + E_{23} X E_{23}^\dagger +E_{34} X E_{34}^\dagger +E_{41} X E_{41}^\dagger ) \nonumber \\ +y \, (E_{13} X E_{13}^\dagger + E_{31} X E_{31}^\dagger +E_{24} X E_{24}^\dagger +E_{42} X E_{42}^\dagger ) +z \, (E_{14} X E_{14}^\dagger + E_{43} X E_{43}^\dagger \nonumber \\ +E_{21} X E_{21}^\dagger +E_{32} X E_{32}^\dagger )+ \sum_{i \neq j} \, G_{ij} X G_{ij}^\dagger - \sum_{i \neq j} \, 
F_{ij} X F_{ij}^\dagger \,,
\label{Eq:Km}
\end{eqnarray} 
where $w,x,y,z \geq 0$ and $X$ is a positive matrix. The mapped matrix $\tilde{X}$ is given as
\begin{eqnarray}
\tilde{X} = \Phi[w,x,y,z] (X) =  \left( 
\begin{array}{cccc}
\tilde{x}_{11} & -x_{12} & -x_{13} & -x_{14}\\ 
-x_{21} & \tilde{x}_{22} & -x_{23} & -x_{24}\\ 
-x_{31} & -x_{32} & \tilde{x}_{33} & -x_{34}\\ 
-x_{41} & -x_{42} & -x_{43} & \tilde{x}_{44}\end{array}
\right) \,,
\label{Eq:Mm4t4}
\end{eqnarray}
where $\tilde{x}_{11} = w \, x_{11}+x \, x_{22}+ y\, x_{33}+z \, x_{44}$, $\tilde{x}_{22} = w \, x_{22}+x \, x_{33}+ y\, x_{44}+z \, x_{11}$,  $\tilde{x}_{33} = w \, x_{33}+x \, x_{44}+ y\, x_{11}+z \, x_{22}$, and $\tilde{x}_{44} = w \, x_{44}+x \, x_{11}+ y\, x_{22}+z \, x_{33}$. The map $\Phi[w,x,y,z]$ will be positive if and only if $\tilde{X}$ is a positive matrix. 


We can study the capability of generalized Choi maps to detect certain type of entangled states. However, first we have to make sure that the mapped matrix \ref{Eq:Mm4t4} is positive. We approach this problem by analyzing the principle minors (PM) of the mapped matrix. There are $15$ such PMs. The first order PMs consist of four diagonal elements $\tilde{x}_{jj}$, which are indeed positive. There are $6$ second order  PMs, which can be written as
\begin{eqnarray}
\tilde{\mathcal{M}}_{12} = (w^2 + x z) \rho_{11} \rho_{22} - |\rho_{12}|^2 + \text{14 positive terms} \, ,
\end{eqnarray}
\begin{eqnarray}
\tilde{\mathcal{M}}_{13} = (w^2 + y^2) \rho_{11} \rho_{33} - |\rho_{13}|^2 + \text{14 positive terms} \, ,
\end{eqnarray}
\begin{eqnarray}
\tilde{\mathcal{M}}_{14} = (w^2 + x z) \rho_{11} \rho_{44} - |\rho_{14}|^2 + \text{14 positive terms} \, ,
\end{eqnarray}
\begin{eqnarray}
\tilde{\mathcal{M}}_{23} = (w^2 + x z) \rho_{22} \rho_{33} - |\rho_{23}|^2 + \text{14 positive terms} \, ,
\end{eqnarray}
\begin{eqnarray}
\tilde{\mathcal{M}}_{24} = (w^2 + y^2) \rho_{22} \rho_{44} - |\rho_{24}|^2 + \text{14 positive terms} \, ,
\end{eqnarray}
\begin{eqnarray}
\tilde{\mathcal{M}}_{34} = (w^2 + x z) \rho_{33} \rho_{44} - |\rho_{34}|^2 + \text{14 positive terms} \, .
\end{eqnarray}
All $6$ PMs will be positive either for $w \geq 1$ or ($y \geq 1$ and $x z \geq 1$). There are $4$ third order $\tilde{\mathcal{M}}_{ijk}$ PMs, and we have analyzed that for all of them, the principle minors have the following possible negative terms, which will be positive for $ w \geq 1$ (10 terms), $y \geq 1$ (1 term), and $ x z \geq 1$ (1 term). We also have $2$ negative terms without parameters $(w,x,y,z)$ and $52$ positive terms. The condition $x z \geq 1$ is not a must provided $w \geq 1$ as seen before (see supplimentary material). The final principle minor $\tilde{\mathcal{M}}_{1234}$ have the following possible negative terms, which will be positive for $w \geq 1$ (60 terms), $y \geq 1$ (26 terms), and $x z \geq 1$ (9 terms). There are $8$ other negative terms and $154$ positive terms. Quite remarkably the conditions for positivity are summarized as
\begin{eqnarray}
w \geq 1\,, \quad y \geq 1, \, \quad x z \geq 1 \,. 
\end{eqnarray}
If $w < 1$, then the other two conditions must be met, however further analysis needs to done to explore that part. These conditions merely indicate that if all of them are satisfied simultaneously, the map is positive. The first condition $w\geq 1$ ensure expected positivity as most of possible negative terms are dependent on it, among the two others, $x z \geq 1$ is not necessary as the possible negative terms related with it are the lowest. In fact, $3$ positive maps provided in \cite{Hou-JPA43} are $\Phi_1(2,1,0,0)$, $\Phi_2(2,0,1,0)$, and $\Phi_1(2,0,0,1)$. The product $x z = 0$ for all of these special maps, $w\geq 1$ for all $3$ of them and $y \geq 1$ for only 1 map. So we conclude that $w \geq 1$ is must whereas $y \geq 1$ may or may not hold, depending on value assumed by $w$. As the possible negative terms dependent on $w$ are more than other terms, taking a larger value of $w$ ensures positivity. We prove the positivity of our proposed maps as follows. The positive maps defined in \cite{Hou-JPA43} are subset of our proposed maps. So we claim that maps $\Phi[2, x, 0, 0]$, $\Phi[2, x, 1, 0]$, and $\Phi[2, x, 0, 1]$ are indeed positive maps for $x\geq 1$. The proof is quite simple and based on the fact that all the terms related with parameter $x$ in all of the principle minors are strictly positive terms and hence taking any non-zero value of $x$ maintain the positivity of these maps. Below, we show the detection power of these maps.    

\section{$4 \otimes \, 4$ system}
\label{Sec:M}
Let us consider a quantum state  defined as 
\begin{eqnarray}
\rho_{\beta, \gamma} = \frac{\beta}{13+\gamma} \, \sigma_1 + \frac{\gamma}{13+\gamma} \, \sigma_2 + \frac{10-\beta}{13+\gamma}\, \sigma_3 + \frac{3}{13+\gamma} \, \sigma_4 \,,
\end{eqnarray}
where 
\begin{eqnarray}
\sigma_1 = \frac{1}{4}  \, (|01\rangle\langle 01| + |12\rangle\langle 12|+|23\rangle\langle 23|+|30\rangle\langle 30|) \,,
\end{eqnarray}
\begin{eqnarray}
\sigma_2 = \frac{1}{4}  \, (|02\rangle\langle 02| + |13\rangle\langle 13|+|20\rangle\langle 20|+|31\rangle\langle 31|) \,,
\end{eqnarray}
\begin{eqnarray}
\sigma_3 = \frac{1}{4}  \, (|03\rangle\langle 03| + |10\rangle\langle 10|+|21\rangle\langle 21|+|32\rangle\langle 32|) \,,
\end{eqnarray}
\begin{eqnarray}
\sigma_4 = | \psi \rangle\langle \psi| \,,
\end{eqnarray}
\begin{eqnarray}
|\psi \rangle = \frac{1}{2} \, (|00\rangle + |11\rangle + |22\rangle + |33\rangle)\,,
\end{eqnarray}
and $0 \leq \beta \leq 10$, $\gamma \geq 0$. The entanglement properties on this state have been studied in \cite{Jafarizadeh-EPJD55}. The state $\rho_{\beta, \gamma}$ is PPT for $1 \leq \beta \leq 9$ and $\gamma \geq 3$, NPT for $0 \leq \beta < 1$, $\beta > 9$ or $\gamma < 3$. The state is PPT entangled (bound entangled) for $1 \leq \beta < 3$, $7 < \beta \leq 9$ and $\gamma \geq 3$. For the range $3 \leq \beta \leq 7$ and $\gamma > 3$, the entanglement properties are not known. The only possible negative eigenvalue of the state $\tilde{\rho}_{\beta, \gamma} = \mathcal{I}_4 \otimes \Phi (\rho_{\beta, \gamma})$ is given as
\begin{eqnarray}
\lambda = \frac{-9+3 w + (10-\beta) z + x \beta + y \gamma}{52 + 4 \gamma} \,.
\end{eqnarray}

For positive map $\Phi[2,1,0,0]$, $\lambda$ is negative (hence states are entangled) for $0 \leq \beta < 3$, therefore this map is not completely positive as it detects some NPT as well as PPT entangled states. However, it fails to detect NPT and PPT-entangled part for $7 < \beta \leq 10$. As the map $\Phi[2,x,0,0]$ is positive for $x = 1$ and due to fact that any positive value of $x$ introduces only positive terms in principle minors, therefore we conclude that the map $\Phi[2,x,0,0]$ is a positive map for $x \geq 1$. In this case, $\lambda = (-3 + x \, \beta)/(52+4 \gamma)$. This value is negative as long as $x \, \beta < 3$, e.g. If $x = 1.5$, then $\lambda$ is negative for $0 \leq \beta < 2$, hence some NPT part is detected for $0 \leq \beta < 1$ as well as some PPT-entangled part for $1 \leq \beta < 2$.     
 
The positive map $\Phi[2, 0, 1, 0]$ leads to $\lambda = (-3 + \gamma)/(52+4\gamma)$, which is negative for $\gamma < 3$, therefore this map detects all NPT entangled states, whereas fails to detect PPT-entangled states. The positive map $\Phi[2, x, 1, 0]$ leads to $\lambda = (-3 + \gamma + x \, \beta)/(52+4\gamma)$, which can be negative for $\gamma + x \, \beta < 3$.

The third map $\Phi[2,0,0,1]$ leads to $\lambda = (7-\beta)/(52+4\gamma)$, which means that map is able to detect bound entangled state $7 < \beta \leq 9$ and also for NPT states $\beta > 9$. As our analysis verifies that non-zero values of parameter $x$ only introduce positive terms in all of the principle minors, therefore we conclude that all such maps $\Phi[2, x, 0, 1]$ are positive maps for any positive value of parameter $x$. The $\lambda = (7 - \beta + x \beta)/(52+4\gamma)$. This eigenvalue can be negative for $x < 1$, e.g. for $x = 0.1$, $\lambda$ is negative for $7.778 < \beta \leq 10$. Hence maps are able to detect PPT-entangled states as well as NPT part.  

\section{$2 \otimes 4$ systems}
\label{Sec:N}

In this section, we aim to check the entangleemnt detection power of generalized Choi maps for $2 \otimes 4$ quantum systems. For the most general quantum state $\rho$ in $2 \otimes 4$ system, we observe that $\tilde{\rho} = \mathcal{I}_2 \otimes \Phi[w,x,y,z] (\rho)$ gives us
\begin{eqnarray}
\tilde{\rho} =  \left( 
\begin{array}{cccccccc}
\tilde{\rho}_{11} & -\rho_{12} & -\rho_{13} & -\rho_{14} & \tilde{\rho}_{15} & -\rho_{16} & -\rho_{17} & -\rho_{18}\\ 
-\rho_{21} & \tilde{\rho}_{22} & -\rho_{23} & -\rho_{24} & -\rho_{25} & \tilde{\rho}_{16} & -\rho_{27} & -\rho_{28} \\ 
-\rho_{31} & -\rho_{32} & \tilde{\rho}_{33} & -\rho_{34} & -\rho_{35} & -\rho_{36} & \tilde{\rho}_{37} & -\rho_{38}\\ 
-\rho_{41} & -\rho_{42} & -\rho_{43} & \tilde{\rho}_{44} & -\rho_{45} & -\rho_{46} & -\rho_{47} & \tilde{\rho}_{48}\\
\tilde{\rho}_{51} & -\rho_{52} & -\rho_{53} & -\rho_{54} & \tilde{\rho}_{55} & -\rho_{56} & -\rho_{57} & -\rho_{58}\\
-\rho_{61} & \tilde{\rho}_{62} & -\rho_{63} & -\rho_{64} & -\rho_{65} & \tilde{\rho}_{66} & -\rho_{67} & -\rho_{68}\\
-\rho_{71} & -\rho_{72} & \tilde{\rho}_{73} & -\rho_{74} & -\rho_{75} & -\rho_{76} & \tilde{\rho}_{77} & -\rho_{78}\\
-\rho_{81} & -\rho_{82} & -\rho_{83} & \tilde{\rho}_{84} & -\rho_{85} & -\rho_{86} & -\rho_{87} & \tilde{\rho}_{88}
\end{array}
\right) \,,
\label{Eq:Mm8t8}
\end{eqnarray}
where we have defined 
\begin{eqnarray}
\tilde{\rho}_{11} = w \, \rho_{11}+x \, \rho_{22}+y\, \rho_{33}+z\, \rho_{44}\,, \nonumber \\
\tilde{\rho}_{22} = w\, \rho_{22}+x \, \rho_{33}+z\,\rho_{11}+y\, \rho_{44} \,,\nonumber \\
\tilde{\rho}_{33} = w\, \rho_{33}+z\, \rho_{22}+y\,\rho_{11}+x\,\rho_{44} \,,\nonumber \\
\tilde{\rho}_{44} = w\, \rho_{44}+y\,\rho_{22}+z\,\rho_{33}+x\,\rho_{11} \,,\nonumber \\
\tilde{\rho}_{55} = w \, \rho_{55}+x\,\rho_{66}+y\,\rho_{77}+z\,\rho_{88} \,,\nonumber \\
\tilde{\rho}_{66} = w\,\rho_{66}+z\,\rho_{55}+x\,\rho_{77}+y\,\rho_{88} \,,\nonumber \\
\tilde{\rho}_{77} = w \, \rho_{77}+y\,\rho_{55}+z\,\rho_{66}+x\,\rho_{88} \,,\nonumber \\
\tilde{\rho}_{88} = w \, \rho_{88}+x\,\rho_{55}+y\,\rho_{66}+z\,\rho_{77} \,,\nonumber \\
\tilde{\rho}_{15} = w \, \rho_{15}+x\,\rho_{26}+y\, \rho_{37}+z\,\rho_{48} \,,\nonumber \\
\tilde{\rho}_{26} = w \, \rho_{26}+z\,\rho_{15}+x\, \rho_{37}+y\,\rho_{48} \,,\nonumber \\
\tilde{\rho}_{37} = w \, \rho_{37}+y\,\rho_{15}+z\,\rho_{26}+x\,\rho_{48} \,,\nonumber \\
\tilde{\rho}_{48} = w \, \rho_{48}+x\,\rho_{15}+y\,\rho_{26}+z\,\rho_{37} \,.\nonumber \\
\label{Eq:mem}
\end{eqnarray}
All other off-diagonal matrix elements get only negative sign. It is interesting to observe that the last four matrix elements defined in Eq.(\ref{Eq:mem}) are dependent on each other only. This clearly indicates which types of entangled states may be detected by generalized Choi maps and which states simply can not be detected. We will provide one example for each bound entangled states as well as NPT entangled state. 

The well known bound entangled states for $2 \otimes 4$ are given \cite{Horodecki-PLA232} as
\begin{eqnarray}
\sigma_b = \frac{1}{7 b+1} \, \left[ 
\begin{array}{cccccccc}
b & 0 & 0 & 0 & 0 & b & 0 & 0 \\ 
0 & b & 0 & 0 & 0 & 0 & b & 0\\ 
0 & 0 & b & 0 & 0 & 0 & 0 & b\\ 
0 & 0 & 0 & b & 0 & 0 & 0 & 0\\
0 & 0 & 0 & 0 & \frac{1+b}{2} & 0 & 0 & \frac{\sqrt{1-b^2}}{2}\\
b & 0 & 0 & 0 & 0 & b & 0 & 0\\
0 & b & 0 & 0 & 0 & 0 & b & 0\\
0 & 0 & b & 0 & \frac{\sqrt{1-b^2}}{2} & 0 & 0 & \frac{1+b}{2}
\end{array}
\right] \,,
\label{Eq:bes1}
\end{eqnarray}
where $0 < b < 1$. The partial transpose of this matrix w.r.t A or B is positive (PPT), however $\sigma_b$ is entangled for $0 < b < 1$\cite{Horodecki-PLA232}. As the initial off-diagonal matrix elements (last four elements defined in Eq.(\ref{Eq:mem})), responsible for bringing the parameters $(w,x,y,z)$ in off-diagonal positions, are all zero, so the only places where $(w,x,y,z)$ may apear are among diagonal elements. The mapped matrix may have some negative eigenvalues for these parameters, provided that the non-zero off-diagonal matrix elements may become larger than the corresponding diagonal entries. The matrix form of the transformed matrix is given as
\begin{eqnarray}
\mathcal{I}_2 \otimes \Phi (\sigma_b) = \frac{1}{7 b+1} \, \left[ 
\begin{array}{cccccccc}
b \, f & 0 & 0 & 0 & 0 & -b & 0 & 0 \\ 
0 & b \, f & 0 & 0 & 0 & 0 & -b & 0\\ 
0 & 0 & b \, f & 0 & 0 & 0 & 0 & -b\\ 
0 & 0 & 0 & b \, f & 0 & 0 & 0 & 0\\
0 & 0 & 0 & 0 & g & 0 & 0 & -\frac{\sqrt{1-b^2}}{2}\\
-b & 0 & 0 & 0 & 0 & h & 0 & 0\\
0 & -b & 0 & 0 & 0 & 0 & i & 0\\
0 & 0 & -b & 0 & -\frac{\sqrt{1-b^2}}{2} & 0 & 0 & j
\end{array}
\right] \,,
\label{Eq:bes2}
\end{eqnarray}
where we have defined $f = w+x+y+z$, $g = (w + z + b \, (w + 2 \, x + 2\, y +z))/2$, $h = (z+y + b(2 \, w + 2 \, x + y + z))/2$, $i = (x+y + b(2 \, w + x + y + 2\, z))/2$, and $j = (w+x + b(w + x + 2\, y + 2\,z))/2$. As we have argued before that $w$ must be $1$ or larger, therefore, it is obvious that $f > 1$ for all the acceptable choices of parameters. In addition, all the diagonal elements are strictly larger than off-diagonal elements, therefore it is not possible for these maps to detect this family of bound entangled states. In our recent work, we had claimed that reduction map (a special case of generalized Choi maps) can detect such states \cite{Ali-QIP24}. During preparation of this work, I found some missing terms in Eq.(\ref{Eq:Km}) in the computer programm, which lead to this erroneous result. All other results and conclusions in that work \cite{Ali-QIP24} are correct except this particular $2 \otimes 4$ bound entangled example. One of the aim of this section is to clarify this confusion as well. We stress that in fact, we should be looking for a map $\Lambda$ from $\mathcal{C}^ 4$ to $\mathcal{C}^ 2$ such that $\mathcal{I}_2 \otimes \Lambda (\sigma_b) < 0$, however, it is known that finding such maps is not an easy task \cite{Horodecki-PLA232}, and despite some efforts, we are still unable to find such maps. Therefore, we conclude that generalized Choi maps and their variants can not detect these particular bound entangled states. However, they may detect some other PPT entangled states in $2 \otimes 4$ systems. It would be an interesting result to seach whether there is such family of bound entangled states in this dimension of Hilbert space. 

It is worth to mention here that the local unitary operations can not change the entanglement properties of a state. We replace one of the pure entangled state, used in the construction of these bound entangled states, with another pure entangled state and find the equivalent quantum states in the same computational basis as 
\begin{eqnarray}
\varrho_b = \frac{1}{7 b+1} \, \left[ 
\begin{array}{cccccccc}
b & 0 & 0 & 0 & 0 & 0 & 0 & -b \\ 
0 & b & 0 & 0 & b & 0 & 0 & 0\\ 
0 & 0 & b & 0 & 0 & 0 & 0 & 0\\ 
0 & 0 & 0 & b & 0 & 0 & b & 0\\
0 & b & 0 & 0 & b  & 0 & 0 & 0\\
0 & 0 & 0 & 0 & 0 & \frac{1+b}{2} & \frac{\sqrt{1-b^2}}{2} & 0\\
0 & 0 & 0 & b & 0 & \frac{\sqrt{1-b^2}}{2} & \frac{1+b}{2} & 0\\
-b & 0 & 0 & 0 & 0 & 0 & 0 & b
\end{array}
\right] \,.
\label{Eq:bes2}
\end{eqnarray}
We have checked all $64$ possibilities of local unitaries which can shuffle the matrix elements, to obtain all such bound entangled states, which 'look' different in the same computational basis, however, they are equivalent and share same entanglement structure. We find that all $'64'$ such states are not at all effected (detected) by $\Phi[w,x,y,z]$ maps. We have checked $64$ local unitaries constructed from Pauli matrices $\sigma_j \otimes \sigma_x \otimes \sigma_j$ with $j = 1, x, y, z$ wihout loss of generality.

\section{Discussion and Summary} 
\label{Sec:Conc}

In this work, we have presented a systematic extension of generalized Choi positive maps to the four-dimensional matrix algebra $M_4(\mathcal{C})$. We rigorously established the positivity of the constructed maps for most general positive matrices and demonstrated their effectiveness in detecting bound entangled states. The identified states are positive under partial transposition for certain range of parameter and NPT for some other. We clearly demonstrate the detection capability of our proposed positive maps to provide explicit example of PPT entanglement in $4\otimes 4$ systems. We have studied the detection capabilities of these maps for 
$2 \otimes 4$ systems and demonstrated analytically that such maps can not detect entanglement of well known bound entangled states. 

Our results contribute to the broader understanding of entanglement detection in higher dimensions, where the structure of the PPT region remains poorly characterized. The generalized Choi Kye maps enrich the existing toolbox of positive-map–based criteria and illustrate that indecomposability can be preserved under controlled dimensional generalization. Beyond their immediate application to PPT entanglement detection, the maps introduced here may also find use in constructing new classes of entanglement witnesses and in exploring the geometry of quantum state spaces in high-dimensional systems.
Although the present work focuses on the theoretical construction of positive (but not completely positive) maps for entanglement detection, the proposed approach can in principle be connected to experimentally measurable quantities. Through the Jamiołkowski–Choi isomorphism, every positive map considered in this work corresponds to an entanglement witness operator W. The expectation value $Tr (W \, \rho)$ can be evaluated experimentally by decomposing W into a linear combination of locally measurable observables, such as tensor products of Pauli operators or local projective measurements. Such decompositions are routinely implemented in current experimental platforms including photonic entanglement setups, trapped-ion systems, superconducting qubits, and nuclear magnetic resonance simulators. In these platforms, the density matrix of multipartite states can be reconstructed via quantum state tomography or partially probed through witness-based measurements, allowing direct experimental verification of entanglement detection schemes based on positive maps. Several experimental realizations of entanglement witnesses and related measurements have already been reported in the literature, demonstrating the feasibility of implementing similar detection strategies in realistic quantum systems \cite{Guehne-PR474, Eisert-RMP82}.
Future work may focus on generalizing these constructions to arbitrary dimensions, analyzing their extremality and optimality properties, and identifying potential operational roles of the corresponding bound entangled states in quantum information processing tasks such as secret sharing or activation of distillability. 
%
\subsection*{Author Contributions}
I declare that this is my own work, thought, worked out and compiled all by myself.

\subsection*{Funding}
Not Applicable

\subsection*{Data Availability}
The author can provide any data and material upon request to relevant party.

\subsection*{Code Availability}
The author can provide source code used to compile the results upon request to relevant party.

\section*{Declarations}

\subsection*{Conflict of Interest}
The author declare that they have no conflict of interest.

\subsection*{Ethics approval}
Not Applicable

\subsection*{Consent to participate}
Not Applicable

\subsection*{Consent to publications}
Not Applicable

\end{document}